\documentclass[useAMS]{mn2e}
\usepackage{times,epsfig}

\def\lsim{\,\lower2truept\hbox{${<\atop\hbox{\raise4truept\hbox{$\sim$}}}$}\,}
\def\gsim{\,\lower2truept\hbox{${> \atop\hbox{\raise4truept\hbox{$\sim$}}}$}\,}
\def\simlt{\mathrel{\rlap{\lower 3pt\hbox{$\sim$}}
        \raise 2.0pt\hbox{$<$}}}
\def\simgt{\mathrel{\rlap{\lower 3pt\hbox{$\sim$}}
        \raise 2.0pt\hbox{$>$}}}

\title[The WMAP-NVSS correlation]{A reassessment of the evidence of the Integrated Sachs-Wolfe effect through the WMAP-NVSS correlation}
\author[Raccanelli et al.]{\parbox[t]{\textwidth}{A. Raccanelli$^{1}$, A. Bonaldi$^{2,1}$, M. Negrello$^{3}$, S. Matarrese$^{4}$, G. Tormen$^{1}$ and G. De Zotti$^{2,5}$}
\vspace*{8pt}\ \\
$^{1}$Dipartimento di Astronomia, Universit\`a di Padova, Vicolo dell'Osservatorio 2, I-35122 Padova, Italy \\
$^{2}$INAF, Osservatorio Astronomico di Padova, Vicolo dell'Osservatorio 5, I-35122 Padova, Italy \\
$^{3}$The Open University, Walton Hall, Milton Keynes MK7 6AA, UK\\
$^{4}$Dipartimento di Fisica ``Galileo Galilei'', Universit\`a di Padova, and INFN Sezione di
Padova, via F. Marzolo, 8 I-35131 Padova, Italy\\
$^{5}$SISSA, Via Beirut 4, I-34014, Trieste, Italy}
\date{}
\begin{document}
\maketitle
\begin{abstract}

We reassess the estimate of the cross-correlation of the spatial distribution of the NRAO VLA Sky Survey (NVSS) radio sources with that of Cosmic Microwave Background (CMB) anisotropies from the Wilkinson Microwave Anisotropy Probe (WMAP). This re-analysis is motivated by the fact that most previous studies adopted a redshift distribution of NVSS sources inconsistent with recent data. We find that the constraints on the bias-weighted redshift distribution, $b(z)\times {\cal N}(z)$, of NVSS sources, set by the observed angular correlation function, $w(\theta)$, or, equivalently, by the power spectrum of their spatial distribution, strongly mitigate the effect of the choice of ${\cal N}(z)$. If such constraints are met, even highly discrepant redshift distributions yield NVSS--WMAP cross-correlation functions consistent with each other within statistical errors. The models favoured by recent data imply a bias factor, $b(z)$, decreasing with increasing $z$, rather than constant, as assumed by most previous analyses. As a consequence, the function $b(z)\times {\cal N}(z)$ has more weight at $z<1$, i.e. in the redshift range yielding the maximum contribution to the ISW in a standard $\Lambda$CDM cosmology. On the whole, the NVSS turns out to be better suited for ISW studies than generally believed, even in the absence of an observational determination of the redshift distribution. The systematics introducing spurious power in the angular correlation function on scales of several degrees are strongly reduced restricting the analysis to the sub-sample brighter than 10 mJy. Even though this sub-sample comprises less than one third of the NVSS sources, it yields a slightly more significant detection of the ISW effect than the full sample ($3\sigma$ rather than $2.5\sigma$). The NVSS--WMAP cross-correlation function is found to be fully consistent with the prediction of the standard $\Lambda$CDM cosmology.

\end{abstract}

\begin{keywords}
cosmic microwave background --- cosmological parameters --- cosmology: observations --- radio continuum: galaxies.
\end{keywords}

\section{Introduction}

As first pointed out by Crittenden \& Turok (1996) a promising way of probing the (linear) Integrated Sachs \& Wolfe (1967) (ISW) effect is through correlations of Cosmic Microwave Background (CMB) maps with tracers of large scale structure. Since the ISW effect shows up on large angular scales, a significant application of this test had to await until the first all-sky high quality CMB temperature maps have been provided by WMAP (Bennett et al. 2003; Hinshaw et al. 2007). The WMAP data have been cross correlated with a variety of radio, IR, optical, and X-ray surveys (see Aghanim et al. 2007 for a review) to look for evidences of a decay of the gravitational potential due to the influence of dark energy.

The first (marginally) significant correlations (at the $\simeq 2.5\sigma$ level) of the WMAP first year data were found by Boughn \& Crittenden (2004, 2005) with the HEAO-1 A2 full sky hard X-ray map (Boldt 1987)  and with a map of the number density distribution of 1.4 GHz radio sources detected by the NRAO VLA Sky Survey (NVSS), covering about 80\% of the sky. The cross correlation between NVSS and WMAP 1st year data was confirmed by Nolta et al. (2004), and by Vielva et al. (2006) using different techniques, in real, harmonic and wavelet spaces. Further analysis was triggered by the release of the WMAP 3 year data. Some groups exploited again the NVSS data (Pietrobon et al. 2006; McEwen et al. 2007, 2008).

Independently of the analysis technique used, the comparison of the correlations inferred from the data with model predictions require two basic ingredients: the redshift distribution and the bias parameter of sources. Since redshift measurements are available only for a tiny fraction of NVSS sources, it is necessary to resort to model estimates. All the investigations quoted above basically adopted the redshift distribution given by model RLF1 of Dunlop \& Peacock (1990) that was shown by  Boughn \& Crittenden (2002) to reproduce the autocorrelation function of NVSS sources assuming a redshift-independent bias parameter (see also Cress \& Kamionkowsky 1998, Magliocchetti et al. 1999). However, as pointed out by Blake et al. (2004) and Negrello et al. (2006), this model overestimates the density of low-redshift radio sources (Magliocchetti et al. 2000), which was found to be consistent with Dunlop \& Peacock's (1990) pure luminosity evolution model (Magliocchetti et al. 2002). But if the adopted redshift distribution, and the associated bias model, are incorrect, the conclusions of the analysis may be flawed. In fact, the contribution to ISW signal in a $\Lambda$CDM cosmology peaks at $z\simeq 0.4$ and rapidly declines at higher and lower redshifts (Afshordi 2004).

This situation has motivated our re-visitation of the problem. In \S\,2 we discuss the interpretation of the autocorrelation function of NVSS sources in the light of the results by Negrello et al. (2006). In \S\,3 we provide a new estimate of the WMAP--NVSS cross-correlation as a function of the angular separation and of the associated errors using both the Internal Linear Combination (ILC) temperature map built using the 3 year WMAP data (Hinshaw et al. 2007) and the foreground cleaned CMB map obtained by Bonaldi et al. (2007) using the same data and allowing for the presence of an ``anomalous'' dust-correlated Galactic emission, in addition to the canonical synchrotron, free-free and thermal dust emissions. In \S\,4 the observed cross-correlation function (CCF) is compared with models prediction. The main conclusions are presented and discussed in \S\,5.

\section{The two-point angular correlation function of NVSS sources}

The NVSS (Condon et al. 1998) is the largest-area radio survey at $1.4\,$GHz. It covers $\sim10.3\,$sr of the sky north of $\delta=-40^{\circ}$; the source catalogue contains 1.8$\times$10$^6$ sources. The two-point angular correlation function, $w(\theta)$, of NVSS
sources has been measured by Blake $\&$ Wall (2002a, 2002b) and Overzier et al. (2003) for different flux-density thresholds between 3 mJy and 500 mJy. The overall shape of $w(\theta)$ is
well reproduced by a double power-law. On scales below $\sim 0.1^\circ$, the steeper power-law reflects the distribution of the resolved components of single giant radio sources. On larger
scales the shallower power-law describes the correlation between distinct radio sources.  The latter provides insights on the way in which they trace the underlying dark matter distribution, and on the cosmological framework which determines the distribution of dark matter at each epoch. In the present context we are interested only on the large-scale behaviour.

Earlier investigations of the ISW effect used NVSS sources down to the formal detection limit of $\simeq 2.5\,$mJy, although the completeness is only $\simeq 50\%$ at the faintest fluxes, while it rapidly increases with increasing flux, reaching 99\% at 3.4 mJy (Condon et al. 1998). Also, systematic surface density gradients, yielding spurious contributions to $w(\theta)$ (or to the power spectrum), are approximately negligible only for $S_{1.4\rm GHz}\geq 10$ mJy (Blake \& Wall 2002a). Following Negrello et al. (2006), we will therefore include in our analysis only sources brighter than 10 mJy. The NVSS source surface density at this threshold is $16.9\,\hbox{deg}^{-2}$.

The angular correlation function of a population of extragalactic sources is related to their spatial correlation function, $\xi(r,z)$, and to their redshift distribution, ${\mathcal N}(z)$, by Limber's (1953) equation:
\begin{equation}
w(\theta)=\int dz{\mathcal N}^{2}(z)\int d(\delta z)
\xi[r(\delta z,\theta),z]\left[\int dz{\mathcal N}(z)\right]^{-2}.
\label{eq:wth_obj}
\end{equation}
Here, $r(\delta z,\theta)$ is the comoving spatial distance between two objects located at redshifts $z$ and $z+\delta z$ and separated by an angle $\theta$ on the sky.

Equation~(\ref{eq:wth_obj}) shows that the contributions to $w(\theta)$ of a population of extragalactic sources are weighted by the square of their redshift distribution. The ${\mathcal N}(z)$ yielded by Dunlop \& Peacock (1990) model RLF1, adopted in previous ISW studies, has a spike at low $z$, which, as pointed out by Negrello et al. (2006), is inconsistent with the recent accurate determinations of local luminosity functions of NVSS sources (Magliocchetti et al. 2002; Sadler et al. 2002; Mauch \& Sadler 2007). Also, Ho et al. (2008) found that such redshift distribution is inconsistent with the results of cross-correlations of NVSS with 2MASS and SDSS samples.  As stressed by Magliocchetti et al. (1999), because of the ${\mathcal N}^2(z)$ weighting of the projected spatial correlation function, the spike makes a significant contribution to $w(\theta)$, even though the visibility volume of the corresponding sources is small. Actually, such sources, that have been interpreted as the spiral and starburst galaxy populations emerging at mJy levels, would provide the dominant contribution to the clustering signal on angular scales larger than $\sim 2^\circ$ (Cress \& Kamionkowski 1998; Magliocchetti et al. 1999; Blake et al. 2004). But the contribution of spiral and starburst galaxies to counts at $S_{1.4\rm GHz}\geq 10\,$mJy, computed using the recent determinations of their luminosity function, is very small ($\simeq 0.5\%$; Negrello et al. 2006). Yet, a statistically significant positive $w(\theta)$ was measured on scales $\ge 5^\circ$, and the interpretation of this data proved to be challenging.

On scales large enough for the clustering signal be due to galaxies residing in distinct dark matter haloes and under the assumption of a one-to-one correspondence between sources and their host haloes, the spatial two-point correlation function can be written as the product of the correlation function of dark matter, $\xi_{\rm DM}$, times the square of the bias parameter, $b$ (Matarrese et al. 1997; Moscardini et al. 1998):
\begin{eqnarray}
\xi(r,z)=b^{2}(M_{\rm eff},z)\xi_{\rm DM}(r,z)\ ,
\label{eq:xi_obj}
\end{eqnarray}
where $M_{\rm eff}$ represents the effective mass of the dark matter haloes in which the sources reside. The observationally determined spatial correlation length of star-forming galaxies, $r_0(z= 0)\sim  3$--$4\,\hbox{Mpc}\,\hbox{h}^{-1}$ (Saunders et al. 1992; Wilman et al. 2003) is consistent with an effective halo mass not exceeding $M^{\rm SF}_{\rm eff}= 10^{11} M_\odot \,\hbox{h}^{-1}$. Together with the small contribution of these sources to the counts for $S_{1.4\rm GHz}\geq 10\,$mJy, this implies that they are negligible contributors to the observed $w(\theta)$ (see Negrello et al. 2006 for more details).

The clustering  properties of low-$z$ AGN-fuelled radio sources are much stronger and imply that they reside in dark matter haloes more massive than $\simeq 10^{13.4}\, M_\odot$ (Magliocchetti et al. 2004). These could then easily account for the observed $w(\theta)$ on scales $\lsim 1^\circ$--$2^\circ$. However, a generic prediction of the Cold Dark Matter paradigm of structure formation, is that the spatial correlation function of matter displays a sharp cut-off around a
comoving radius of a few hundred Mpc, and becomes negative on larger scales. For typical redshifts $z\sim 1$, this translates in a cut off of $w(\theta)$ on scales $\sim 2^\circ$. On larger scales, the negative contribution of $z\sim 1$ sources overwhelms the positive contributions from low-$z$ sources {\it if $M_{\rm eff}$ is redshift-independent, as indicated by data on optically selected quasars} (Porciani et al. 2004; Croom et al. 2005). To account for the data in the framework of the standard hierarchical clustering scenario it is necessary that the effective bias factor decreases with increasing $z$, rather than strongly increase like for optical quasars. This adds weight to the contribution of low-$z$ NVSS sources to the matter density fluctuations, enhancing the expected CMB--NVSS cross-correlation.

\begin{figure*}
\begin{center}
\includegraphics[height=6cm, width=8.5cm]{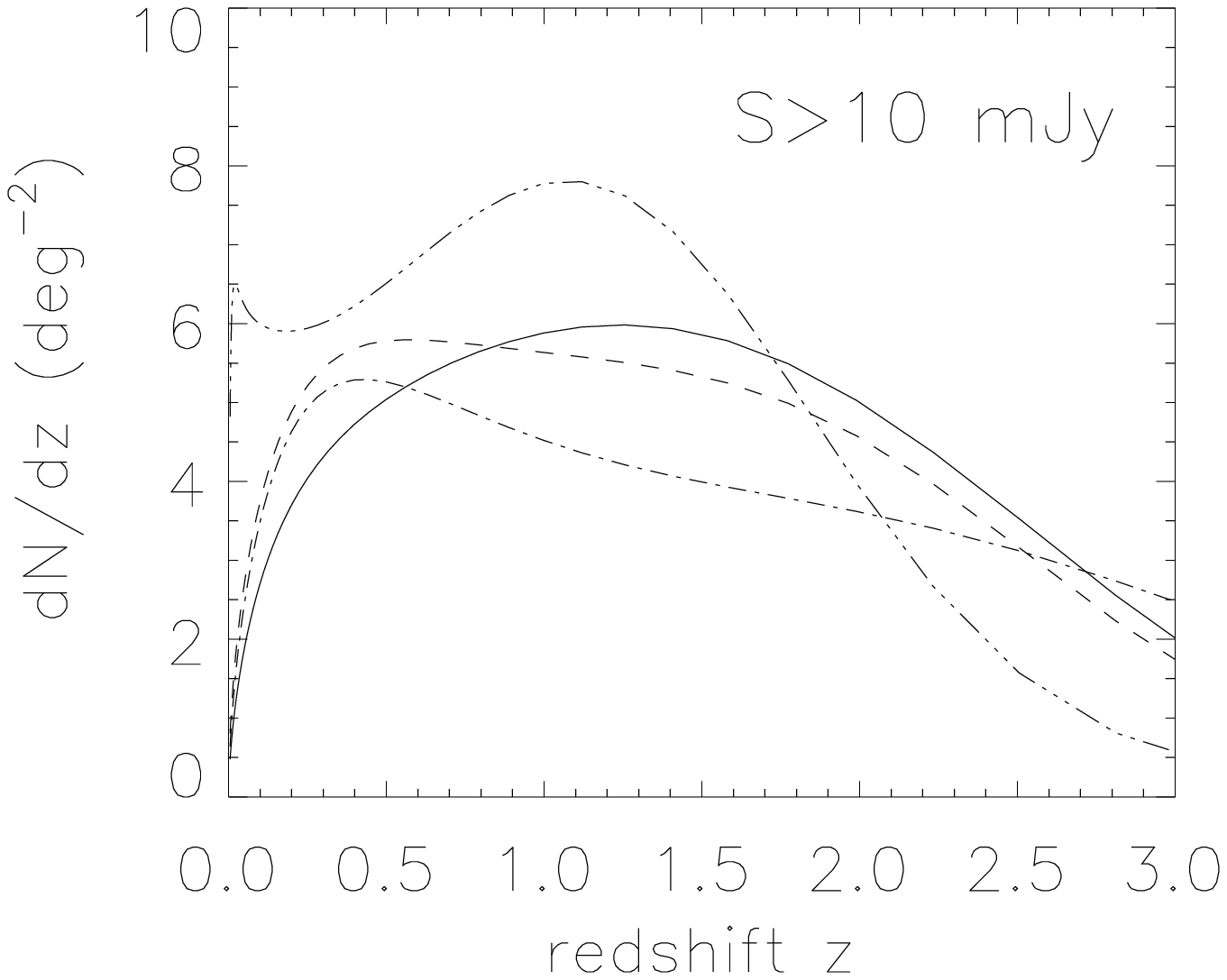}
\includegraphics[height=6cm, width=8.5cm]{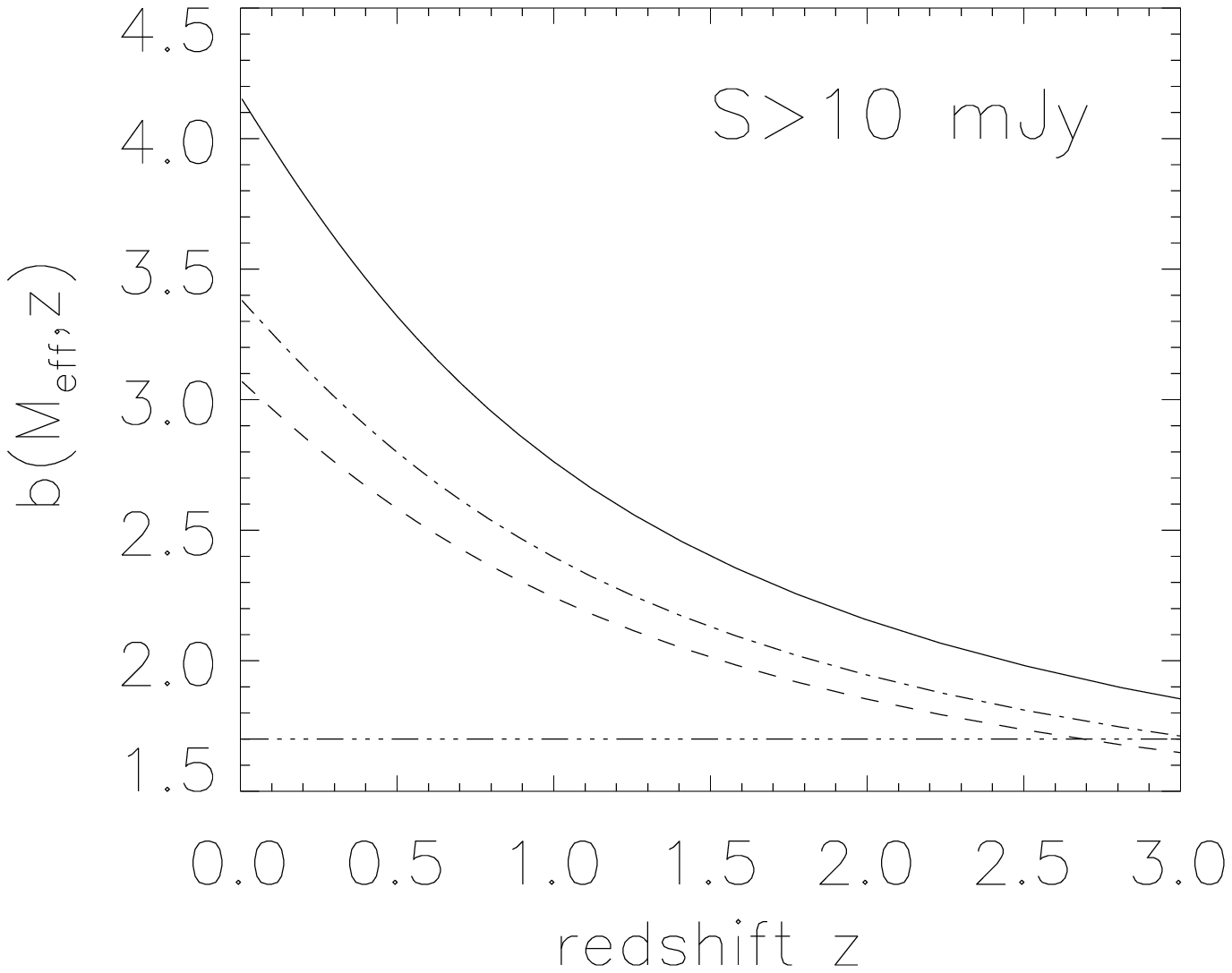}
\includegraphics[height=6cm, width=8.5cm]{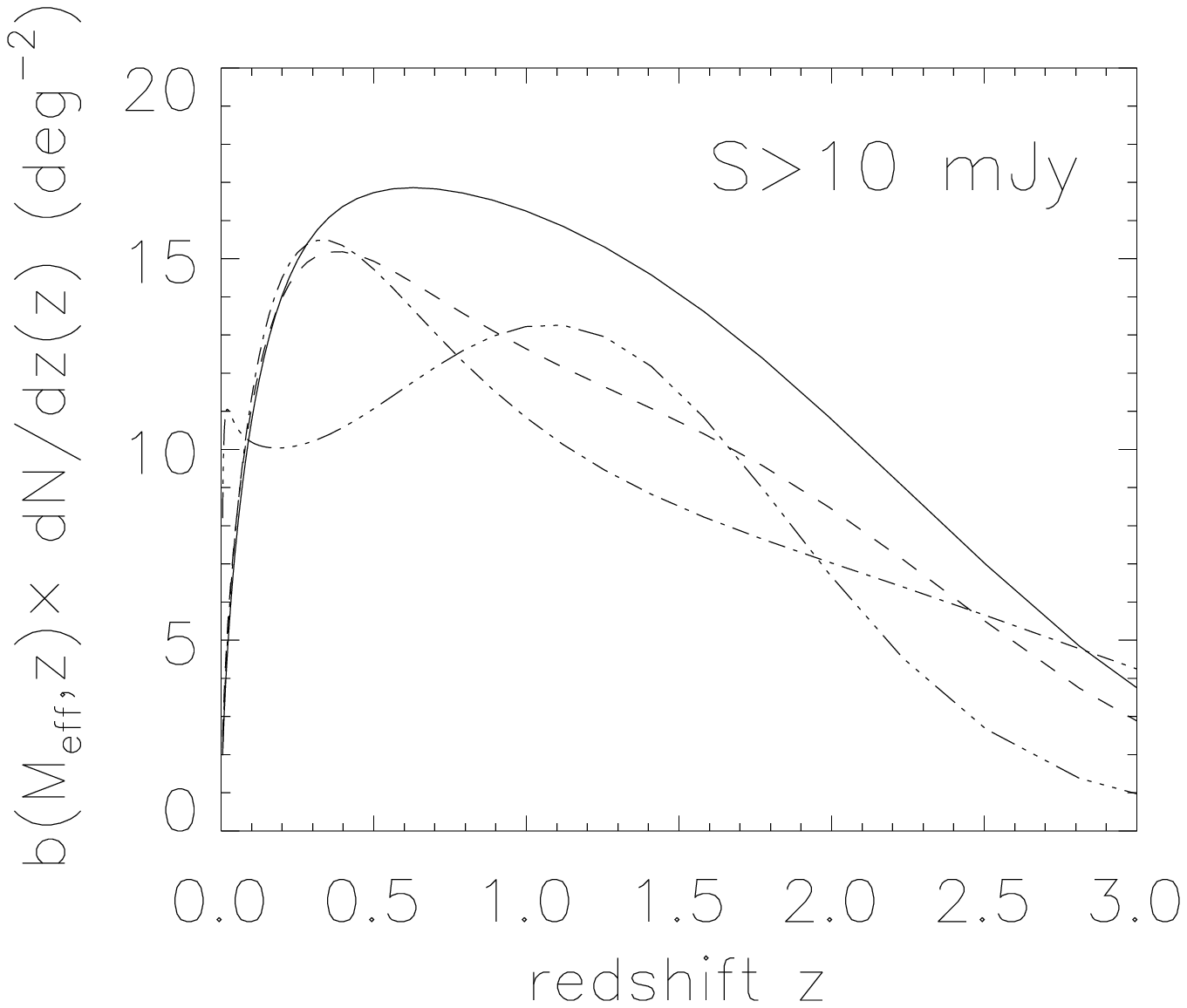}
\includegraphics[height=6cm, width=8.5cm]{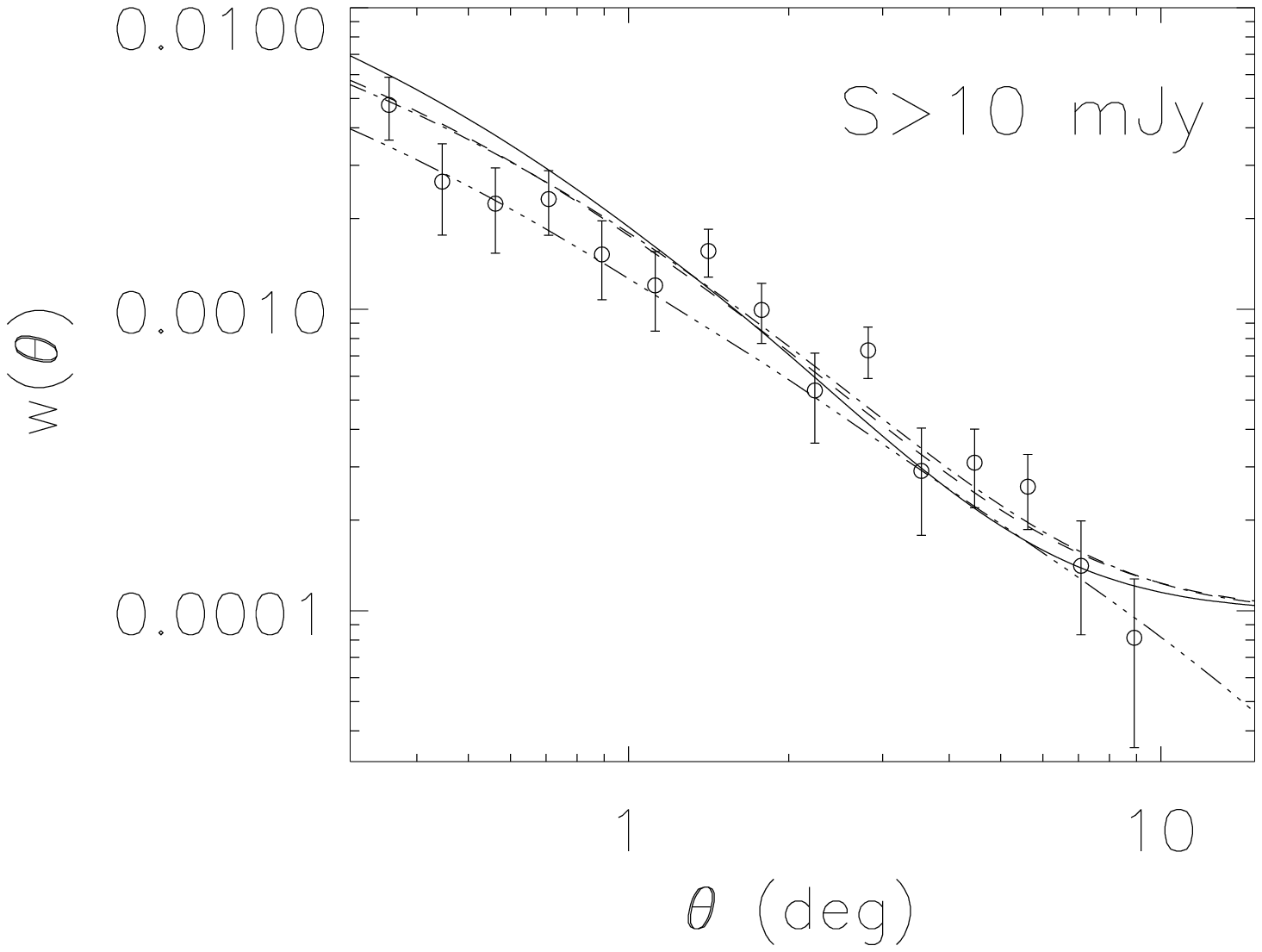}
\caption{Redshift distributions (upper left-hand panel), evolution of the bias factor (upper right-hand panel), bias-weighted redshift distribution (lower left-hand panel), and fits of the observed angular correlation function (lower right-hand panel) of NVSS sources brighter than 10 mJy for Dunlop \& Peacock (1990) pure luminosity evolution (PLE) model (solid line) and the 2 versions of luminosity/density evolution (LDE) model, MEAN-$z$ (dashed line) and HIGH-$z$ (dot-dashed line). The three model $w(\theta)$ include the small systematic offset $\delta w(\theta) = 10^{-4}$ attributed to small systematic variations in the source surface density due to calibration problems (Blake \& Wall 2002a).  The three dots-dashed lines show, for comparison, the results obtained with Dunlop \& Peacock (1990) RFL1 model with a constant value of the bias parameter ($b=1.7$), as assumed by most previous analyses. The data points with error bars show the Blake \& Wall (2002b) estimates of $w(\theta)$. }\label{zdist}
\end{center}
\end{figure*}

Negrello et al. (2006) found that the observed $w(\theta)$ can be reproduced if $M_{\rm eff}$ is proportional to the characteristic mass of virialized systems, $M_{\star}(z)$, with $M_{\rm eff}(z=0) \simeq 10^{15}\,M_\odot$. This implies $M_{\rm eff}(z=1.5) \simeq 10^{13}\,M_\odot$,  close to the value found for optical quasars. A possible interpretation of this result is that AGN-powered radio sources are so rare locally that the typical halo hosting one of them must include a very large number of galaxies, i.e. must correspond to a rich cluster of galaxies. The strong evolution of radio sources implies that they become relatively more abundant with increasing redshift, and the typical halo mass decreases correspondingly. This happens in such a way that the clustering of radio sources reflects that of the largest haloes that collapse at any cosmic epoch.

We have repeated the analysis using the 2 versions (MEAN-$z$ and HIGH-$z$) of the luminosity/density evolution (LDE) model by Dunlop \& Peacock (1990), which are also consistent with the recent data on the local luminosity function. As illustrated by Fig.~\ref{zdist}, the requirement that the observed $w(\theta)$ is reproduced (lower left-hand panel) forces (lower right-hand panel) the bias-weighted redshift distributions, $b(z)\times {\mathcal N}(z)$, corresponding to the 3 models to be very close to each other in the redshift interval from which most of the contribution to the ISW signal comes, thus lessening the effect of uncertainties in the redshift distribution and on the bias factor (upper panels). Also the redshift dependence of the bias factor adds weight to the interesting redshift range, implying that the NVSS sample is very well suited for ISW studies.

\section{Observational estimate of the WMAP--NVSS cross-correlation function}

The empirical cross-correlation function was computed as:
\begin{equation}
\hat{C}^{NT}(\theta_k)=\frac{\sum N_iT_jw_i^Nw_j^T}{\sum w_i^Nw_j^T} \label{emp_ccf},
\end{equation}
where $N_i$ is the number of NVSS sources in the i-th pixel of the map of number count fluctuations, $T_j$ is the temperature fluctuation in the j-th pixel of the CMB map, the two pixels being at an angular distance $\theta_k$. The weights $w_j^T$ and $w_i^N$ are equal to unity for valid pixels or equal to zero if the pixels fall within a region not covered by the survey or masked; the apex $T$ and $N$ refer to the CMB and to the NVSS maps, respectively. The maps are pixelized with the Healpix (G\'orski et al. 2005) resolution parameter NSIDE=32, corresponding to a pixel size of about $1^\circ .8 \times 1^\circ .8$.

We used two CMB maps: the WMAP three-year Internal Linear Combination (ILC) map (Hinshaw et al. 2007) and the foreground-cleaned map yielded by model {\bf M2} of Bonaldi et al. (2007). As for the NVSS map, our choice of a 10 mJy flux limit has dispensed us with the need of introducing corrections for the spurious density gradients in the survey noted by Boughn \& Crittenden (2002) for the sample limited to 2.5 mJy.  We have removed the strip at $|b|\le 5^\circ$ where extragalactic sources are blurred by bright Galactic sources and by small scale structure of the synchrotron and free-free emission. Including the unobserved region at $\delta <-40^\circ$, the NVSS mask covers 25\% of the sky.

The CMB mask adopted is based on the Kp0 mask (Bennett et al. 2003), which excludes the regions most contaminated by Galactic foregrounds and the brightest point sources in the WMAP bands. The Kp0 mask has NSIDE=512, corresponding to a pixel size of about $7'$. We degraded this map to the lower resolution adopted for this work and we set $w_j^T=0$ for all pixels including at least one masked $7'$ pixel. The resulting mask excludes 37\% of the sky. The union of NVSS and CMB masks covers 48\% of the sky.

\begin{figure}
\begin{center}
\includegraphics[totalheight=9cm, width=6cm, angle=90.]{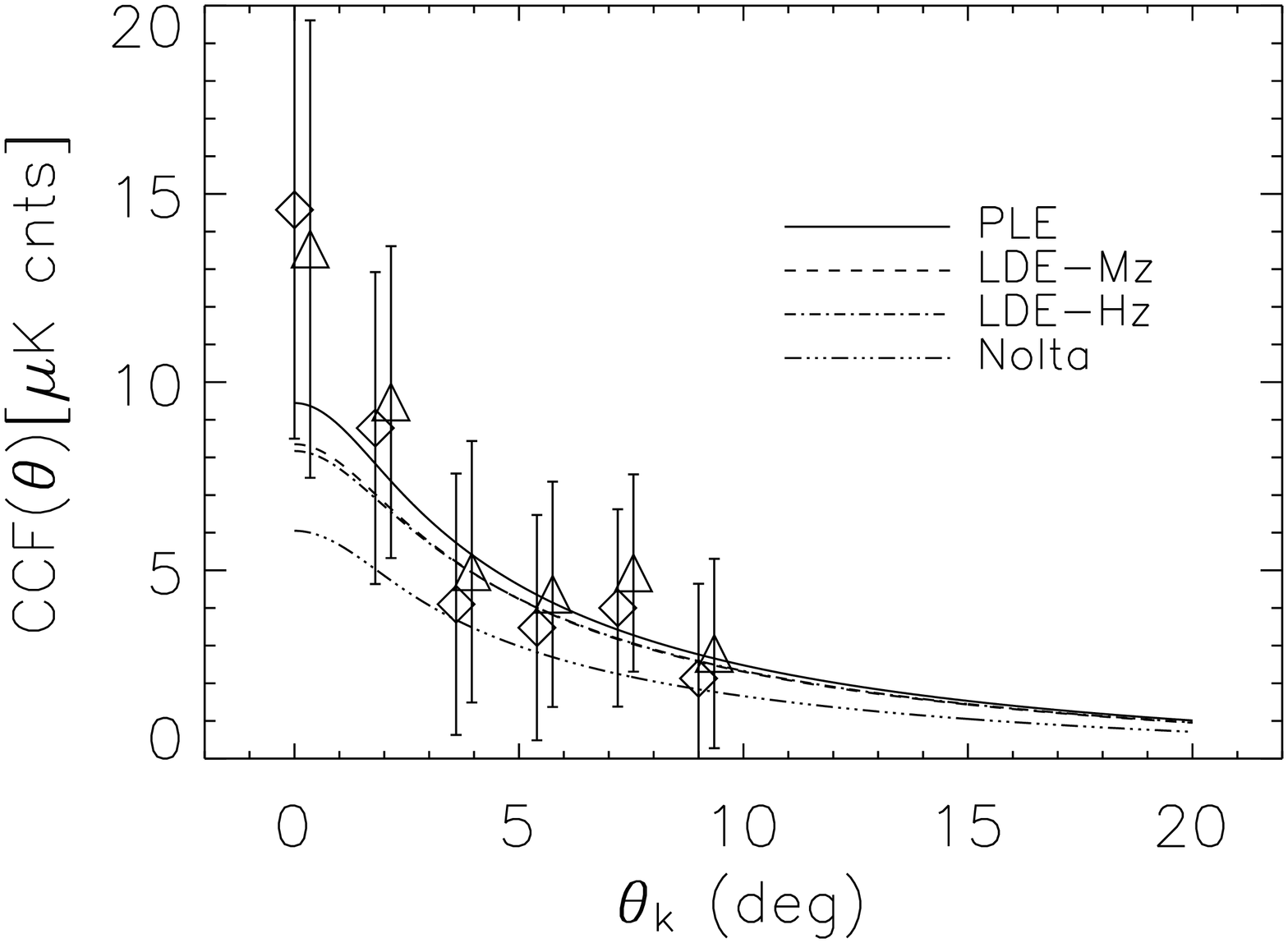}
\caption{Estimates of the WMAP--NVSS cross-correlation function (CCF) using the ILC map (triangles) and the foreground cleaned map by Bonaldi et al. (2007; diamonds), with $1\sigma$ error bars. The ILC results are slightly shifted to the right to improve the readability. The empirical CCF is compared with expectations from the best-fit WMAP 3-year cosmology for the models described in the text. As in Fig.~\protect\ref{zdist}, the solid line refers to the PLE model, the dashed line to the LDE MEAN-$z$ model, the dot-dashed line to the LDE HIGH-$z$ model, and the three dots-dashed line to the model used by Nolta et al. (2004).  } \label{fig:CCF}
\end{center}
\end{figure}

For both CMB maps, we computed the empirical CCF [eq.~(\ref{emp_ccf})] for a set of angular separations, in degrees: $\theta_k=\{0.0,1.8,3.6,5.4,7.2,9.0,10.8\}$. As discussed by Blake \& Wall (2002a) the tiny angular correlations among NVSS sources on scales $\gsim 5^\circ$ [$w(\theta)\sim 10^{-4}$] may well be due to small systematic variations in the source surface density due to calibration problems. This may also add some spurious power to the CCF; the effect is however small and the empirical CCF is indeed found to be consistent with zero on these scales. For each $\theta_k$ we considered pixel pairs whose centers have separations in the range  $\theta_k-\Delta\theta <\theta< \theta_k+\Delta\theta$ with $\Delta\theta=1^\circ$. To estimate the errors we simulated 1000 mock NVSS maps by randomly redistributing the unmasked pixels of the true NVSS map. For each mock NVSS map we computed the CCF with the CMB map. The error on $\hbox{CCF}(\theta_k)$ is then estimated as the rms value of the CCFs for the mock maps. Alternatively, the significance of the signal can be estimated from the fraction of mock CCFs at or above its amplitude. We find that the significance estimated in this way is in excellent agreement with that expected from the S/N ratio for a Gaussian distribution. The results are shown in Fig.~\ref{fig:CCF}. The significance of the  measured signal is similar for the two CMB maps used. The CCF differs from zero at the 2.1--$2.4\sigma$ level for each of the first 2 points, and the overall significance of the ISW signal is 99.7\% or slightly higher, corresponding to $3\sigma$ or somewhat more. Boughn \& Crittenden (2003) reported a somewhat less significant ($2.5\sigma$) detection of the NVSS-CMB cross-correlation, suggesting that the systematics affecting the distribution of the faintest NVSS sources overcome the statistical advantage of their far larger number.



\section{The theoretical cross-correlation function}

Following Nolta et al. (2004) we write the cross-correlation power spectrum between the surface density fluctuations of NVSS sources and CMB fluctuations as:
\begin{equation}\label{eq:CNT}
C_l^{NT} = \langle a_{lm}^N a_{lm}^{T*} \rangle = 4 \pi
\int_{k_{\rm min}}^{k_{\rm max}} \frac{dk}{k} \Delta^2(k) f_l^N(k)
f_l^T(k),
\end{equation}
where $f_l^N$ and $f_l^T$ are the NVSS and CMB filter functions, respectively, and  $\Delta^2(k)$ is the logarithmic matter power spectrum today as a function of the wave number $k$:
\begin{equation}
\Delta_{\delta}^2(k) =  {k^3\over 2 \pi^2} P_{\delta}(k) = A\delta_H^2  \left({ck\over H_0}\right)^{3+n} T_f^2(k).
\end{equation}
For $\delta_H$ we have used the expression given by eq.~(A3) of Eisenstein \& Hu (1998), valid for a flat universe with a cosmological constant. The reference values of cosmological parameters are those given in Table 2 of Spergel et al. (2007), namely $H_0=100h=72\,\hbox{km}\,\hbox{s}^{-1}\,\hbox{Mpc}^{-1}$, $\Omega_m=0.13h^{-2}$, $\Omega_\Lambda=1-\Omega_m$, $n=0.96$, and $\sigma_8=0.78$.  The constant $A$ was computed  normalizing the power spectrum to $\sigma_8=0.78$; we find $A=1.06$. 
The matter transfer function, $T_f(k)$, was computed using CMBfast (Seljak \& Zaldarriaga 1996).

The NVSS filter function writes:
\begin{equation}\label{eq:flN}
f_l^N(k) = \int dz \frac{dN}{dz} b_r(z) {D(z)\over D(0)} j_l(kc\eta(z)),
\end{equation}
where  $(dN/dz)dz$ is the mean number of sources per steradian with redshift $z$ within $dz$, brighter than the flux limit, $b_r(z)$ is the bias factor relating the source overdensity to the mass overdensity, assumed to be scale-independent, $D(z)$ is the linear growth factor of mass fluctuations, $j_l(x)$ is the spherical Bessel function, and $\eta(z)$ is the conformal look-back time:
\begin{equation}
\eta(z) = \int_0^z \frac{dz'}{H(z)} = \int_0^z \frac{dz'}{H_0E(z') },
\end{equation}
with $E(z)=\sqrt{\Omega_{\Lambda} +\Omega_{R}(1+z)^2 + \Omega_{m}(1+z)^3}$, $\Omega_{R}$ being the curvature parameter, set to zero in this paper. For the linear growth factor, appropriate for the large scales of interest here, we adopt the approximation (Carroll et al. 1992):
\begin{eqnarray}
D(z) & = & \frac{5\Omega_m(z)}{2(1+z)} \left\{ \Omega_m(z)^{4/7} -
\Omega_{\Lambda}(z) + \left[ 1+ \frac{\Omega_m(z)}{2} \right] \cdot \right. \nonumber \\
&\cdot & \left. \left[1+ \frac{\Omega_{\Lambda}(z)}{70} \right] \right\}^{-1},
\end{eqnarray}
with $\Omega_m(z) = \Omega_{m}(1+z)^3/E(z)^2$ and $\Omega_{\Lambda}(z) = \Omega_{\Lambda}/E(z)^2$.

As mentioned in \S\,1, we restrict ourselves to NVSS sources brighter than $10\,$mJy and adopt the redshift distributions given by three Dunlop \& Peacock (1990) evolution models for AGN powered sources. As for star-forming galaxies, we estimate the redshift distribution using the local luminosity function of Magliocchetti et al. (2002) and no evolution: evolutionary effects, if any, are irrelevant for our sample since star-forming galaxies brighter of 10 mJy are found to be at $z\lsim 0.1$. The filter function for the ISW effect is:
\begin{eqnarray}\label{eq:flT}
f_l^T(k) & = &3 T_{\rm CMB} \Omega_{m} \left(\frac{H_0}{ck}\right)^2 \cdot \nonumber \\
&\cdot&\int dz \frac{d[D(z)(1+z)/D(0)]}{dz} j_l(k\eta(z)).
\end{eqnarray}
Since eq.~(\ref{eq:flT}) holds for linear perturbations, i.e. for $k\ll 1$, the integration over $k$  of eq.~(\ref{eq:CNT}) requires some caution. We have checked for a set of multipoles $l$ that the cross-correlation function obtained using the power spectrum of CMB temperature perturbation given by CMBfast are well approximated by setting in eq.~(\ref{eq:CNT}) $k_{\rm max} = 0.02\,\hbox{Mpc}^{-1}$.

The cross-correlation function as a function of the angular separation $\theta$ is then obtained as:
\begin{equation}
C^{NT}(\theta) = \sum_{l} \frac{2l+1}{4 \pi}
C_l^{NT}P_l(\cos \theta).
\end{equation}
where $P_l$ are the Legendre polynomials. For the range of scales of interest here, it is enough to sum over the interval $2\le l \le 200$.

Figure~\ref{fig:CCF} shows that, once the constraints from the observed $w(\theta)$ are taken into account, the predicted CCF for the WMAP 3-year values of cosmological parameters is only weakly affected by the uncertainties on the redshift distributions of NVSS sources. We find a good agreement between the predicted and empirical CCFs, somewhat at odds with the finding by Ho et al. (2008) that the WMAP 3-year model predicts an ISW amplitude about $2\sigma$ below their estimate, which exploits various galaxy surveys in addition to the NVSS. The strong constraints from the angular correlation function of the latter sources are not taken into account, however.

\section{Discussion and conclusions}

The NVSS sample is of primary importance for the investigation of the ISW effect because of its very large area (82\% of the sky), its large number of sources (almost $2 \times 10^6$, its depth in redshift, and its immunity to systematics related to the uncertain corrections for dust extinction. It has however drawbacks, not all of which were properly dealt with by previous analyses. On one side, spurious contributions to the power spectrum of the NVSS source distribution is introduced by known systematic effects such as small systematic variations in the source surface density due to calibration problems at low flux densities and to striping effects, particularly at negative declinations (Blake \& Wall 2002a; Boughn \& Crittenden 2002). To minimize this problem, we have restricted our analysis to sources with $S_{1.4 \rm GHz}\ge 10\,$mJy (Blake \& Wall 2002a,b); at this flux limit, striping effects are negligible.

The main potential problem, however, stems from the very limited direct redshift information, forcing us to resort to models. Most previous analyses (Boughn \& Crittenden 2002, 2004; Nolta et al. 2004; Pietrobon et al. 2006; Vielva et al. 2006; McEwen et al. 2007) relied on model RLF1 of Dunlop \& Peacock (1990), which was found by Boughn \& Crittenden (2002) to reproduce the NVSS autocorrelation function. However, as pointed out by Negrello et al. (2006), this model is inconsistent with the recent determinations of the luminosity function of low-$z$ NVSS sources (Magliocchetti et al. 2002; Sadler et al. 2002; Mauch \& Sadler 2007). A totally independent confirmation of this conclusion has been obtained by Ho et al. (2008) by cross-correlating the NVSS with 2MASS and SDSS samples, whose redshift distributions are known. In this way these authors obtained constraints on the bias-weighted redshift distribution, $b(z)\times {\cal N}(z)$, which is the relevant quantity for predicting the NVSS-CMB cross-correlation induced by the ISW effect [eq.~(\ref{eq:flN})].

The large spread of redshift-dependent radio luminosity functions yielded by the original set of Dunlop \& Peacock (1990) models is substantially narrowed down by the more recent data on the radio luminosity function at low redshifts, that are fully consistent with the pure luminosity evolution and by the two luminosity/density evolution models. Strong constraints on $b(z)\times {\cal N}(z)$ follow from the observational determination of the angular correlation function, $w(\theta)$, of NVSS sources, which depends on $[b(z)\times {\cal N}(z)]^2$. In fact, the functions $b(z)\times {\cal N}(z)$ for three models considered here are found to be substantially closer to each other, particularly at low-redshifts, than the corresponding redshift distributions. The decrease of the bias factor with increasing $z$ implied by observational determinations of $w(\theta)$ adds weight to the redshift range yielding most of the contribution to the ISW signal, making the NVSS sample even better suited to test the effects of dark energy on the growth of structure than one would guess from the estimated redshift distribution.

A concern expressed in some previous papers is that the microwave emission of radio sources themselves contribute to the observed WMAP--NVSS correlations. However, the radio source contribution to temperature fluctuations is dominated by the brightest sources, while the contribution to the source number density fluctuations, relevant for studies of the ISW effect is dominated by the faintest sources. Number density and surface brightness fluctuations are thus largely uncorrelated, implying that the source microwave emission rather weakens the correlations. We find that, if no sources are masked, the CCF estimates decrease by 6--10\%. On the other hand, removing just the WMAP sources (Hinshaw et al. 2007; Lopez-Caniego et al. 2007) is enough to make the contamination of the ISW signal negligibly small, consistent with previous results (e.g. McEwen et al. 2007; Ho et al. 2008).

In summary, the main results of this paper are:

\begin{itemize}

\item the constraints on the bias-weighted redshift distribution, $b(z)\times {\cal N}(z)$, of NVSS sources, set by the observed angular correlation function or, equivalently, by the power spectrum of their spatial distribution, strongly mitigate the effect of the large uncertainties on ${\cal N}(z)$. Even highly discrepant redshift distributions, including instances ruled out by recent data, yield NVSS--WMAP cross-correlation functions consistent with each other within statistical errors, although the use of the Dunlop \& Peacock (1990) RLF1 model leads to a slight underestimate of the predicted ISW signal for the WMAP 3 year values of the cosmological parameters. This fact largely alleviates what was perceived as the main weakness of the exploitation of the NVSS for studies of the ISW.

\item The models favoured by recent data imply bias functions, $b(z)$, decreasing with increasing $z$, rather than constant, as assumed by most previous analyses. As a consequence, the function $b(z)\times {\cal N}(z)$ has more weight at $z<1$, i.e. in the redshift range yielding the maximum contribution to the ISW in a standard $\Lambda$CDM cosmology. Again, this means that the NVSS is better suited for ISW studies than generally believed.

\item The systematics (striping, incompleteness, etc.) introducing spurious power in the angular correlation function on scales of several degrees are strongly reduced restricting the analysis to the sub-sample with a flux limit of 10 mJy. Even though this sub-sample comprises less than one third of the NVSS sources, it yields a slightly more significant detection of the ISW effect ($3\sigma$ rather than $2.5\sigma$). The microwave emission of NVSS sources weakens the correlations with WMAP fluctuations, rather than increasing them as previously argued. The effect is, however, negligibly small.

\item The NVSS--WMAP cross-correlation function is found to be fully consistent with the prediction of the standard $\Lambda$CDM cosmology.

\end{itemize}

\section*{Acknowledgements}

We acknowledge useful suggestions from A. Balbi and M. Liguori. Work supported in part by ASI (contracts Planck LFI Phase E2 Activity and COFIS) and MUR.


\end{document}